# Evaluation of molecular orbital symmetry via oxygen-induced charge transfer quenching at a metal-organic interface

Iulia Cojocariu,[1] Henning Maximilian Sturmeit,[2] Giovanni Zamborlini,[2,*] Albano Cossaro,[3] Alberto Verdini,[3] Luca Floreano,[3] Enrico D'Incecco,[4] Matus Stredansky,[3,4] Erik Vesselli,[3,4] Matteo Jugovac,[1] Mirko Cinchetti,[2] Vitaliy Feyer[1,5,*] and Claus M. Schneider[1,5]

[1] *Peter Grünberg Institute (PGI-6), Forschungszentrum Jülich GmbH, Jülich, Germany*

[2] *Technische Universität Dortmund, Experimentelle Physik VI, 44227 Dortmund, Germany.*

[3] *CNR-IOM, Lab. TASC, s.s. 14 km 163,5, 34149 Trieste (Italy)*

[4] *Physics Department, University of Trieste, via A. Valerio 2, 34127 Trieste, Italy*

[5] *Fakultät f. Physik and Center for Nanointegration Duisburg-Essen (CENIDE), Universität Duisburg-Essen, D-47048 Duisburg, Germany*

*Corresponding author: v.feyer@fz-juelich.de, giovanni.zamborlini@tu-dortmund.de

## Abstract

Thin molecular films under model conditions are often exploited as benchmarks and case studies to investigate the electronic and structural changes occurring on the surface of metallic electrodes. Here we show that the modification of a metallic surface induced by oxygen adsorption allows the preservation of the geometry of a molecular adlayer, giving access to the determination of molecular orbital symmetries by means of near-edge x-ray absorption fine structure spectroscopy, NEXAFS. As a prototypical example, we exploited Nickel Tetraphenyl Porphyrin molecules deposited on a bare and on an oxygen pre-covered Cu(100) surface. We find that adsorbed atomic oxygen quenches the charge transfer at the metal-organic interface but, in contrast to a thin film sample, maintains the ordered adsorption geometry of the organic molecules. In this way, it is possible to disentangle π* and σ* symmetry orbitals, hence estimating the relative oscillator strength of core level transitions directly from the experimental data, as well as to evaluate and localize the degree of charge transfer in a coupled system. In particular, we neatly single out the σ* contribution associated with the N 1*s* transition to the mixed N $2p_{x,y}$-Ni $3d_{x^2-y^2}$ orbital, which falls close to the leading π*-symmetry LUMO resonance.

# Introduction

Planar heterocyclic compounds play a crucial role in the development of novel organic-based technologies, such as biosensors, OLEDs, and photovoltaic devices [1]. When contacted with metal surfaces, they frequently self-assemble in 2-dimensional ordered arrays, where the chemical state and the adsorption properties of each molecule are well-defined and homogeneous across the whole surface. At the same time, the molecule-metal interaction results in charge transfer at the interface between the substrate and adsorbed molecules, leading to the filling (emptying) of the lowest unoccupied (highest occupied) molecular orbitals (LUMO/HOMOs) [2–4].

Valence band spectroscopy (VB) and near-edge X-ray absorption fine structure (NEXAFS), together with other absorption-based spectroscopies, are often used to probe the occupation of the LUMOs, as well as the symmetry of the molecular orbitals (MOs) in organic networks. In order to correctly identify the molecular orbitals involved in the charge transfer process and to evaluate the molecule-substrate and intermolecular interactions [5–7], both experimental approaches require reference data for free or weakly interacting molecules. Gas-phase molecules or powder samples, as well as thin molecular films, are commonly employed for this purpose. While the molecules are randomly oriented in both gas phase and powder samples, an ordered multilayer may offer the advantage of probing the molecular orbitals with selective sensitivity to the orbital symmetry by means of NEXAFS measurements at different polarizations. In fact, $\sigma^*$ and $\pi^*$ absorption resonances are selectively measured by exploiting their dependence on the photon beam polarization. The electric field perpendicular (*s*-polarization) or parallel (*p*-polarization) to the incidence plane will yield different intensities (NEXAFS linear dichroism), according to the orientation of the orbitals involved in the induced electronic transitions [8]. In the case of planar aromatic compounds, a flat adsorption geometry yields the best discrimination of the symmetry selected unoccupied MOs. Unfortunately, even in the case of perfectly planar aromatic molecules, a fully coherent growth and a quasi-planar molecular orientation can be hardly achieved beyond the few-layer thickness limit, and only for few specific systems. In particular, epitaxial growth parallel to the surface can be achieved for 2D extended planar molecules like phthalocyanines [3,9–11], or when the substrate lattice offers a good matching with the molecular structure, such as sexi-thiophene on Au(110) [12], Pentacene on Ag(111) [13], Cu(110) [14], $TiO_2$(110) [15]. Neither of these conditions is achieved for heterocyclic molecules that contain substituents with high rotational freedom [16].

Based on these considerations, it becomes evident that it would be extremely convenient to find a reference substrate that preserves, at the same time, the long-range molecular order, the planar orientation and the electronic properties of the free molecule. This reference substrate may be obtained by reducing the molecule-surface interaction. The first step in this direction was to exploit the anisotropic surface corrugation of rutile-$TiO_2$(110) to drive the oriented growth of both polyconjugated aromatics [15,17–21] and porphyrins [22–25]. However, significant charge transfer may take place between large aromatic molecules and defects of the rutile surface [26]. In addition, a strong chemical interaction between the porphyrin tetrapyrrolic center and the titanium dioxide has been reported [27,28]. Alternative approaches for the decoupling of organics from metal surfaces are based either on the intercalation of insulating layers [29] or on the passivation of the surface, e.g., by graphene [30], Sn alloying [31] and oxidation [32,33]. In particular, it has been shown, that for oxygen modified copper surface the covalent nature of the Cu–O interaction yields a strong localization of the surface electrons and inhibits the charge transfer from the metal to the organic overlayer [32,34,35].

In the present communication, we show that the (√2 x 2√2)R45°-oxygen-reconstructed Cu(100) surface, besides quenching the molecule-surface interaction, allows the preservation of an ordered layer, as confirmed by low energy electron diffraction (LEED) and, therefore, can be employed as a template for reference systems to probe both symmetry and occupation of molecular orbitals. As test molecules, we used a tetrapyrrolic compound, namely Nickel Tetraphenyl Porphyrin (NiTPP), and we compared the results of the NiTPP adsorbed on the bare and on the oxygen-reconstructed copper (100) surface by means of valence band photoemission and NEXAFS. We show that, while NiTPP strongly interacts with the bare copper surface leading to the partial filling of the LUMOs, the charge transfer at the interface can be prevented by pre-adsorption of oxygen. Furthermore, the oxygen-induced decoupling preserves the flat adsorption geometry of the porphyrin macrocycle moiety, allowing us to identify the different LUMOs together with their symmetry, in contrast with the case of porphyrin multilayer, where the linear NEXAFS dichroism vanishes due to random molecular orientation.

## Experimental section

The valence band spectra were measured at the NanoESCA beamline of Elettra, the Italian synchrotron radiation facility in Trieste, using an electrostatic photoemission electron microscope (PEEM) set-up described in detail in Ref. [36]. The data were collected with a photon energy of 35 eV and a total energy resolution of 100 meV, using *p*-linearly polarized light. The NEXAFS experiment was performed at the ALOISA beamline, also located at Elettra synchrotron [37]. The spectra across the C and N K-edges were taken in electron yield mode using a channeltron multiplier [37], and they have been further analyzed following the procedure described in Ref. [38]. The orientation of the surface with respect to the linear polarization (*s* and *p*) of the synchrotron beam was changed by rotating the sample around the beam axis, while keeping fixed the incident angle (6° with respect to the surface plane) of the synchrotron light.

The clean Cu(100) surface was prepared by a standard procedure: cycles of $Ar^+$ ion sputtering at 2.0 keV followed by annealing to 800 K. The oxygen-covered Cu(100) surface showing a (√2 x 2√2)R45° reconstruction, as confirmed by low energy electron diffraction (LEED) patterns, was prepared by exposing the Cu(100) surface to 800 L of $O_2$ while keeping it at 500 K [39].

NiTPP molecules (Sigma Aldrich, ≥95% purity) were thermally sublimated at 570 K from a home-made Knudsen cell type evaporator onto the copper substrate kept at room temperature. As the monolayer coverage is long-range ordered, the achievement of the desired coverage was monitored using reflective high-energy electron diffraction (ALOISA) or LEED (NanoESCA).

## Results and discussion

The valence band spectra of NiTPP monolayer deposited on the bare Cu(100) and on the O-Cu(100) surfaces are presented in Figure 1a (top and bottom, respectively). On the bare copper surface, the VB spectrum exhibits two peaks, previously assigned to the gas-phase LUMO/+1 and

LUMO+3. These are two π molecular orbitals that are partially filled, as a result of the strong molecule-metal interaction [4].

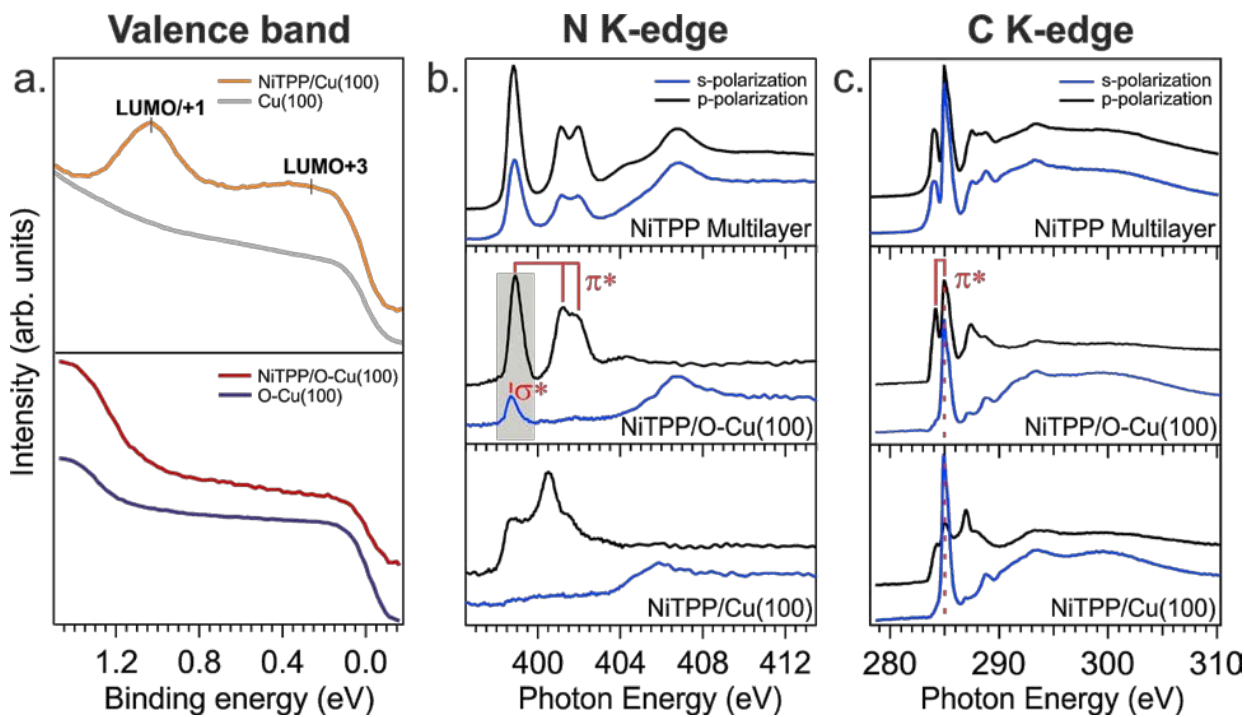

**Figure 1 a)** Valence band spectra of NiTPP on Cu(100) (blue curve) and on O-Cu(100) (black curve); the spectra of bare Cu(100) and O-Cu(100), green and red curve, respectively. **b)** N K-edge NEXAFS spectra of NiTPP multilayer and NiTPP monolayer deposited on bare and oxygen modified Cu(100) surface. **c)** C K-edge NEXAFS spectra of NiTPP multilayer and NiTPP monolayer deposited on bare and oxygen modified Cu(100) surface.

When NiTPP molecules are deposited on the (√2 x 2√2)R45°-oxygen-reconstructed copper surface, the valence band peaks related to the LUMOs features vanish (see Figure 1a, bottom). This observation points towards a significant quenching of the charge transfer at the interface, as a result of the weakening of the interaction between the molecule and the copper substrate favored by the pre-adsorbed oxygen layer. These changes in the electronic structure have dramatic consequences also in the overall appearance of the NEXAFS spectra: while the charge transfer from the substrate to the organic film usually results in the partial or total quenching of the lower-in-energy NEXAFS features, due to the emptying of the starting molecular orbitals involved in the transitions, the quenching of the charge transfer allows the preservation of these features. Moreover, the intensity of the absorption resonances depends on the number of available excitation channels and on the associated oscillator strength of the transitions [40], together with the angle between the electric field of the linearly polarized incoming light and the spatial orientation of the unoccupied MOs.

The nitrogen K-edge spectra of the NiTPP monolayer deposited on bare (bottom) and oxygen modified (middle) copper surface, together with the NiTPP multilayer (top), are compared in Figure 1b. The NEXAFS spectra of NiTPP multilayer on Cu(100) do not show linear dichroism, suggesting a random molecular orientation in the organic layers. The lack of linear dichroism does not allow the determination of the symmetry of the resonances, nor the estimate of the oscillator strength of the transitions. Instead, strong linear dichroism is observed in the N K-edge spectra for the NiTPP monolayer on the bare and on the oxygen pre-covered Cu surfaces. In particular, the resonances dominating the *p*-polarization spectrum below the ionization threshold (~404 eV) display opposite dichroism with respect to the broad resonance recorded in *s*-polarization above the ionization threshold. As a rule of thumb, the N 1s ionization threshold separates the region of π\*-symmetry resonance (below) from that of σ\*-symmetry ones (above). Hence, the observed dichroism points to an adsorption orientation of the porphyrin macrocycle parallel or almost parallel to the surface in both cases.

The flat adsorption geometry of the macrocycle moiety is further confirmed by the dichroic behavior of the C K-edge spectra, both on the bare and on the oxygen pre-covered-Cu(100) substrates (see Figure 1c). Following a well-tested model analysis [5], the NEXAFS resonances at ~284 eV are exclusively associated to the π\*-symmetry LUMO states localized at the macrocycle ring, displaying large dichroism with a very small residual intensity in s-polarization, in full agreement with the observed behavior of the low energy N resonances. Instead, the intense resonance at 284.9 eV is mainly ascribable to the transition to the π\*-symmetry LUMO of the peripheral phenyl rings. The adsorption tilt angle ($\gamma$) of the phenyl rings with respect to the copper surface can be evaluated from the angle-dependent intensity ratios ($Is/Ip$) of the resonance at 284.9 eV, following the equation $\gamma = \frac{1}{2} arctan^2(Is/Ip)$ [38]. Using this approach, tilt angles of 69±5° and 52±5° have been derived for NiTPP on Cu(100) and O-Cu(100), respectively.

The charge transfer occurring at the NiTPP/Cu(100) interface strongly affects the intensity of the π\* resonances and of the σ\* feature as well: all low energy LUMOs are partially filled (and shifted). On the contrary, the close similarity between the resonance intensity and the shape of the multilayer spectra with that of the monolayer grown on the oxidized surface confirms the weak interaction of the latter substrate with the porphyrin array, which now allows us to disentangle the symmetry of the orbitals.

On the oxygen modified system, the low energy features in the N K-edge spectra measured with *p*- and *s*-polarized light are related to π\* (N 1*s*→LUMO/LUMO+1 and LUMO+3) and to σ\* (N 1*s*→LUMO+2) resonances, respectively [41]. The latter transition from N 1*s* to a mixed Ligands $2p_{x,y}$ - Metal $3d_{x^2-y^2}$ orbital falls close to the main π\* symmetry resonance also in porphyrins and phthalocyanines coordinated to other metals, e.g. Cu [38,42] and Co [43], where the precise energy position around the main π\* transition depends on the specific metal element [44] and the local interaction with the substrate [45]. Because of its intrinsically low intensity, this σ\* transition is hardly identified in disordered films [46,47]. Even in the case of homogeneously oriented multilayers, its intensity and energy position can be hardly evaluated, because of the overlap with the residual intensity of the large π\* resonance in s-polarization due to the molecular tilt off the surface. In the present case, the almost perfect parallelism of the porphyrin macrocycle to the surface allows us to perform a quantitative analysis of the lowest π\* and σ\* resonances at 398.95 eV and 398.75 eV, respectively (see Fig.1b, highlighted in grey). It has to be noted that the LUMO+2 associated to σ\* resonance does not yield any clear signature in the VB spectrum, but it is crearly observable in the NEXAFS spectrum. Remarkably, density functional theory calculations showed that the filling of this orbital is responsible for the reduction of the oxidation state of the Ni ion, from Ni(II) to Ni(I) [4,41].

Far above the ionization threshold, the intensity of the partial electron yield results from the photoemission from a core-level into a continuum of states, and it is independent of the polarisation of the impinging electromagnetic radiation. By normalizing to unity the high energy tail of the spectra measured with *s*- and *p*-polarization, we can evaluate the relative oscillator strengths (proportional to the intensity of the resonances) for the aforementioned low energy NEXAFS transitions (and in general, for the resonances below the ionization threshold). After evaluation of the areas of the spectral features, we get a ratio of 4.5 to 1 ($\pi^*$ to $\sigma^*$). It is worth noting that the estimated ratio can be used as an input for the evaluation of the absolute absorption cross-sections in theoretical calculations.

## Conclusions

By means of NEXAFS and valence band spectroscopies, we confirmed that oxygen pre-adsorption on the Cu(100) surface is an efficient approach for suppressing the interaction at the metal-organic interface. Moreover, the O-Cu(100) substrate preserves the planar orientation of deposited NiTPP molecules, allowing us to unambiguously identify the symmetry of frontier molecular orbitals. The corresponding planar monolayer phases are better suited than thin-film NEXAFS measurements to properly model the fine details of the electronic structure, including the molecular orbital symmetry. This method of acquisition of reference spectra can be extended to other absorption based spectroscopies, such as high-resolution electron energy loss spectroscopy (HREELS) and Infrared Reflection Absorption Spectroscopy (IRAS), in the study of adsorption behaviour of heteroaromatic compounds.

## Acknowledgment

This work has partially received funding from the EU-H2020 research and innovation programme under grant agreement No 654360 having benefitted from the access provided by CNR-IOM in Trieste (Italy) within the framework of the NFFA Europe Transnational Access Activity.
M.C., G.Z. and H.S. acknowledge funding from the European Research Council (ERC) under the European Union's Horizon 2020 research and innovation programme (Grant Agreement No. 725767—hyControl). Support from Italian MIUR under project PRIN-2017KFY7XF is also acknowledged.